\documentclass[prl,aps,twocolumn,superscriptaddress,nofootinbib]{revtex4}
\usepackage{mathrsfs}
\usepackage{amsfonts}
\usepackage{amsmath}
\usepackage{amssymb}
\usepackage{array}
\usepackage{verbatim}
\usepackage{epsfig}
\usepackage{graphicx}
\usepackage{amsbsy}
\usepackage{bm}
\usepackage[dvipsnames]{xcolor}  
\usepackage{graphicx} 
\usepackage[font=small,labelfont=bf]{caption} 
\usepackage[subrefformat=parens]{subcaption}

\usepackage{import}
\input{epsf}
\usepackage{psfrag}
\usepackage{xcolor}
\usepackage{slashed}
\usepackage{hyperref}
\usepackage{ulem}

\begin{document}

\begin{flushright}

\end{flushright}

\title{Small-$x$ evolution of dipole amplitude in momentum space: {\it forward} - {\it off-forward} correspondence}

\author{Sanskriti Agrawal}
\affiliation{Department of Physics, Aligarh Muslim University, Aligarh - $202001$, India.}

\author{Raktim Abir}
\affiliation{Department of Physics, Aligarh Muslim University, Aligarh - $202001$, India.}
\email{raktim.ph@amu.ac.in}

\begin{abstract}
We have shown that the small-$x$ evolution of the {\it off-forward} leading-log dipole scattering amplitudes, both pomeron and odderon, in the momentum space can be completely determined by the evolution of the respective {\it forward} amplitudes, 
with rescaled momenta. In position space, if there is translation symmetry (assumption of a large nucleus), the dipole cross section depends on the positions of quarks and anti-quarks only through their separation. The present study is an equivalent proposition in the momentum space - where translation symmetry in momentum bifurcates the amplitudes into two translationally symmetric functions along the ${\bm k}$ line in the ${\bm k}-{\bm \Delta}$ plane. It also shows that high energy evolutions of dipole GTMDs can be achieved only by studying the evolution of dipole TMDs at small-$x$.  
\end{abstract}

\maketitle
{\color{Red} $\bullet$} {\color{NavyBlue} {\it Introduction}:~}
%
The discovery of asymptotic freedom about fifty years ago by David Gross, Frank Wilczek, and David Politzer is considered a triumph in the development of modern particle physics \cite{Gross:1973id,Politzer:1973fx}. The discovery not only made the perturbative results reliable even for the strong interactions, but it firmly established the quantum field theory with the color degree of freedom, $i.e.$, Quantum Chromodynamics (QCD), as the theory of strong interactions. Prior to that, the general disillusion in attempting quantum field theory to strong interaction propelled the development of methods that use the unitarity and analyticity constraints to the scattering amplitude, extending Regge's idea of complex angular momentum mainly by Gribov and his disciples. In the years afterward, two important developments that led to the foundation of the scale and energy evolution of the strong interaction cross-sections are the Dokshitzer-Gribov-Lipatov-Alterelli-Parisi (DGLAP) \cite{Gribov:1972ri,Altarelli:1977zs,Dokshitzer:1977sg} and the Balitsky-Fadin-Kuraev-Lipatov (BFKL) equations \cite{Balitsky:1978ic,Kuraev:1977fs}. While the DGLAP equation sums up higher-order contributions enhanced by the logarithm of virtuality or scales $i.e.$ $\ln Q^2$, the BFKL equation sums the contributions enhanced by the logarithm of high energy, $\ln s$, or by the logarithm of the small momentum fraction, $\ln(1/x)$ - leading to the small-$x$ limit of QCD.
Nevertheless, the resummation of the parameter $\bar{\alpha}_s\ln(1/x)$, in the dilute regime of low parton density, as accomplished by the BFKL evolution equation, is a linear approximation of more complex theory. While it explains the rise of structure-function as one moves to small-$x$ or high energy, the equation ultimately violates the unitarity constraints of the cross-section. The BFKL equation, therefore, does not account for the saturation of the gluons at small-$x$. The initial attempt to solve the unitarity problem was by introducing the non-linear correction to the BFKL equation, which resulted in the Gribov-Levin-Ryskin and Mueller-Qiu (GLR-MQ) evolution equation \cite{Gribov:1983ivg,Mueller:1985wy}. 
It was later shown that the leading $s$-logarithm for high energy scattering can also be obtained from the non-local Wilson line correlators at large $N_c$ limit, both with the operator product expansion and in Mueller dipole model, leading to the Balitsky-Kovchegov (BK) equation \cite{Balitsky:1995ub,Kovchegov:1999yj}. BFKL equation can be viewed as the linearised form of the non-linear BK equation.

%

%
%
%

%
In the study of small-$x$ evolution, the evolution of Wilson line correlators is often the central object to study. These correlators are connected 
to the $S$-matrix of the color dipole interacting with the target. The scattering $S$-matrix, for a quark-anti-quark dipole interacting with a `large' nucleus can be expressed as,   
\begin{eqnarray}
    S({\bm x},{\bm y}) = \frac{1}{N_c}\langle \text{Tr}~ U({\bm x})~U^{\dagger}({\bm y}) \rangle, 
\end{eqnarray}
where ${\bm x}$ and ${\bm y}$ are the positions of quark and anti-quark in the transverse plane, $U({\bm z})=P \exp \left(ig \int z^- \bm A(z^-, {\bm z})\right)$ is a light like Wilson-line in the fundamental representation. The $S$-matrix intrinsically depends on the $x$ variable and can be written in terms of pomeron $N({\bm x},{\bm y})$ and odderon $O({\bm x},{\bm y})$ amplitudes, 
\begin{eqnarray}
    S({\bm x},{\bm y}) = 1-N({\bm x},{\bm y}) + iO({\bm x},{\bm y}). 
\end{eqnarray}
Under charge conjugation, quark and anti-quark swap their positions, and in the eikonal and leading log approximation, this leads to,  
\begin{eqnarray}
    S({\bm x},{\bm y}) = S^{\dagger}({\bm y},{\bm x}). 
\end{eqnarray}
Therefore, under the exchange of positions, $({\bm x} \leftrightarrow {\bm y})$, while odderon $O({\bm y},{\bm x})=-O({\bm x},{\bm y})$ is anti-symmetric, the 
pomeron amplitude $N({\bm x},{\bm y})=N({\bm y},{\bm x})$ is symmetric. 
Imposition of translation symmetry results to the fact that $N({\bm x},{\bm y})$ depends on ${\bm x}$ and ${\bm y}$ only through the transverse separation  
${\bm r} = {\bm x} - {\bm y}$ of the dipole.    
This pointed to an ideal situation when the target size is way larger than the size of the dipole in the transverse plane and the 
impact parameter ${\bm b} = ({\bm x} + {\bm y})/2$ dependence of the pomeron $N$ and odderon $O$ amplitudes can be neglected, 
\begin{eqnarray}
N\left({\bm x},{\bm y}\right) \approx N\left({\bm x}-{\bm y}\right)  ; ~~~ O({\bm x},{\bm y}) \approx O\left({\bm x}-{\bm y}\right). 
\end{eqnarray}

\vspace{0.5cm}
\noindent In this study, we have shown that within a reasonable assumption (certain translation symmetries in the momentum space - which is at par with well-studied translation invariance in the position space), the high-energy evolution of dipole scattering amplitude in the momentum space, with non-zero momentum transfer, can be fully expressed in terms of associated amplitudes in the forward limit with rescaled momenta. In particular, we have shown that translation symmetry in the momentum space translates to the following relations: for pomeron amplitude,
\begin{eqnarray}
\partial_{Y}{\cal N}({\bm k}, {\bm \Delta}) \sim \frac{1}{2} \partial_{Y}  \left[ {\cal N}\left({\bm k} - \frac{{\bm \Delta}}{2},{\bm 0}\right) + {\cal N}\left({\bm k} + \frac{{\bm \Delta}}{2},{\bm 0}\right) \right] \label{pomeron}
\end{eqnarray}
and odderon amplitude in the momentum space,
\begin{eqnarray}
\partial_{Y} {\cal O}({\bm k}, {\bm \Delta}) \sim \frac{1}{2} \partial_{Y} \left[ {\cal O}\left({\bm k} - \frac{{\bm \Delta}}{2},{\bm 0}\right) - {\cal O}\left({\bm k} + \frac{{\bm \Delta}}{2},{\bm 0}\right) \right]   \label{odderon}
\end{eqnarray}
where, $Y=\ln\left(1/x\right)$ and $\bm k$ and $\bm\Delta$ are momentum conjugate to $\bm r$ and $\bm b$ respectively. As a consequence, the small-$x$ evolution of generalised transverse momentum dependent distributions (GTMDs) of gluons, can now be completely expressed in terms of the evolution of two transverse momentum dependent distributions (TMDs). Substantial progress in the area of small-$x$ evolution of TMDs has been made in the last two decades, and a wealth of results are now available on the theory front. However, the GTMD studies are relatively new, and the viability of experimental probes of GTMDs is not that elusive anymore. Being relatively new, GTMD studies are still limited to either model calculations or near-forward limits (the small $\bm \Delta$ approximation). The result presented in this paper may help to move beyond the near-forward approximation. One may now use the known results of TMDs to get the GTMDs for gluons.

\vspace{0.5cm}
{\color{Red} $\bullet$} {\color{NavyBlue} {\it Small-x evolution of pomeron amplitude ${\cal N}({\bm k}, {\bm \Delta})$ and symmetry of kernels:~}}
%
%
In the large $N_c$ limit, the evolution equation for $N({\bm x},{\bm y})$ in the leading log approximation is governed by the Balitsky-Kovchegov (BK) equation,
%
\begin{eqnarray} 
&& \!\!\!\!\!\!\!\! \frac{\partial}{\partial Y} N\left({\bm x},{\bm y}\right)=\frac{{\bar \alpha}_s}{2\pi}\int d^2{\bm z} \frac{\left({\bm x}-{\bm y}\right)^2}{\left({\bm x}-{\bm z}\right)^2\left({\bm z}-{\bm y}\right)^2}\left[N\left({\bm x},{\bm z}\right) \right. ~ \nonumber \\
   && \left. \!\!\!\!\!\!\!\!  ~~~~~~~~~  +N\left({\bm z},{\bm y}\right)-N\left({\bm x},{\bm y}\right)- N\left({\bm x},{\bm z}\right)N\left({\bm z},{\bm y}\right)\right]. \label{BKposition}
\end{eqnarray}
%
This small-$x$ evolution of pomeron amplitude with non-linear unitarity correction was first derived by Balitsky in the framework of high energy effective theory and independently by Kovchegov using the Mueller dipole model \cite{Balitsky:1995ub,Kovchegov:1999ua}. The BK equation resumes BFKL pomeron fan diagrams in large $N_c$ limit. BK equation may also be obtained as the mean-field approximation within the color-glass condensate effective theory. 
This integro-differential equation has two well-known features: 
\begin{itemize}
    \item All four terms on the right side in Eq.\eqref{BKposition} have identical kernel $\omega$,  
    \begin{eqnarray}
     \omega = \frac{\left({\bm x}-{\bm y}\right)^2}{\left({\bm x}-{\bm z}\right)^2\left({\bm z}-{\bm y}\right)^2}.
     \end{eqnarray}
    \item  The kernel along with the measure, $i.e.$, $d^2\bm z ~\omega$ is invariant under the linear 
    fractional transformations or specifically Möbius transformations as, 
\begin{eqnarray}
{\bm z} \rightarrow \frac{\alpha {\bm z} + \beta}{\gamma {\bm z} + \delta},
\end{eqnarray}
(along with identical transformation over ${\bm x}$ and ${\bm y})$ where $\alpha$, $\beta$, $\gamma$ and $\delta$ are complex numbers with $\alpha \delta - \beta \gamma =1$. 
\end{itemize}
Linear fractional transformations are conformal maps as all their generators are conformal: multiplicative inversion $z\rightarrow 1/z$ and affine transformations $z\rightarrow \alpha z + \beta $.
Therefore, the leading order Balitsky-Kovchegov equation has $PSL(2, \mathbb{C})$ symmetry, and extensive studies have been done on this aspect \cite{Gubser:2011qva}. \\

\vspace{0.5 cm}
\noindent  As $PSL(2, \mathbb{C})$ includes $ISO(2)$, or more specifically translation, the equation is invariant under the following translation symmetry, 
\begin{eqnarray}
({\bm x},{\bm z},{\bm y}) \rightarrow \left({\bm x}+{\bm a},{\bm z}+{\bm a},{\bm y}+{\bm a}\right).
\end{eqnarray}
Employing this translation symmetry in position space, the non-linear evolution of the pomeron and odderon in momentum space has been studied \cite{Motyka:2005ep}. 
The assumption of this translation symmetry in position space excludes the momentum transfer $\bm \Delta$ dependence from the beginning.

\vspace{0.5 cm}
\noindent In the following, we will discuss how much symmetry would remain when the equation is transformed to momentum space. The standard way to Fourier transform the dipole amplitudes is the following,   
\begin{eqnarray}
\!\!\!\!\!\!\!\! && \!\!\!\!\!\!\!\!  {\cal N}({\bm k}, {\bm \Delta}) = \int \frac{d^2{\bm x}}{2\pi}\frac{d^2{\bm y}}{2\pi} \nonumber \\ 
&& ~ \times   \exp\left(i{\bm k}({\bm x}-{\bm y})+i {\bm \Delta}
\frac{({\bm x}+{\bm y})}{2}\right)~ \frac{N({\bm x}, {\bm y})}{({\bm x} -{\bm y})^2}.
\end{eqnarray}
\noindent In the momentum space, the equation takes the following form \cite{Hatta:2022bxn}, 
\begin{widetext}
\begin{eqnarray}
\partial_Y {\cal N}({\bm k}, {\bm \Delta})&=& \frac{\bar{\alpha}_s}{\pi} \int \frac{d^2 {\bm k}'}{\left({\bm k} - {\bm k}' \right)^2} \left \{{\cal N}({\bm k}', {\bm \Delta}) -\frac{1}{4} \frac{\left({\bm k}-\frac{{\bm \Delta}}{2}\right)^2}{\left({\bm k}'-\frac{{\bm \Delta}}{2}\right)^2}~{\cal N}({\bm k}, {\bm \Delta})-\frac{1}{4}\frac{\left({\bm k}+\frac{{\bm \Delta}}{2}\right)^2}{\left({\bm k}'+\frac{{\bm \Delta}}{2}\right)^2}~{\cal N}({\bm k}, {\bm \Delta}) \right \} \nonumber \\
&& ~~~~~~~~~~~~~~~~~~ - \frac{\bar{\alpha}_s}{2\pi}\int d^2{\bm \Delta}' {\cal N}\left({\bm k} + \frac{{\bm \Delta}'}{2}, {\bm \Delta} - {\bm \Delta}'\right)
{\cal N}\left({\bm k} + \frac{{\bm \Delta}'-{\bm \Delta}}{2}, {\bm \Delta}'\right), \label{equa01}
\end{eqnarray}
\end{widetext}
%
%
%
%
%
%
\begin{itemize}
    \item In the first term, which we will refer as ${\cal A}_1(\bm k,\bm \Delta)$, within the curly brackets, the pomeron amplitude ${\cal N}({\bm k}', {\bm \Delta})$ is symmetric in ${\bm \Delta} \rightarrow -{\bm \Delta}$ and the kernel does not contain any ${\bm \Delta}$. Therefore, ${\cal A}_1$ is symmetric in ${\bm \Delta} \rightarrow -{\bm \Delta}$. In this term,  
the kernel together with measure $i.e.$, $d^2{\bm k}'\omega_1$, where $\omega_1=1/\left({\bm k} - {\bm k}' \right)^2$, is invariant under the following transformation, 
\begin{eqnarray}
({\bm k}', {\bm k}, {\bm \Delta}) \rightarrow  (\alpha {\bm k}' + \beta, \alpha {\bm k} + \beta, {\bm \Delta}).
\end{eqnarray}
As the kernel does not contain any ${\bm \Delta}$, the momentum transfer ${\bm \Delta}$ is more of a parameter than a variable here during the ${\bm k}'$ integration. 
%
%
 \item Quite interestingly, kernel and measure combination in the second term ${\cal A}_2(\bm k,\bm \Delta)$,  
\begin{eqnarray}
d^2{\bm k}'\omega_2  = d^2 {\bm k}'\frac{\left({\bm k} - \frac{\bm \Delta}{2}\right)^2}{\left({\bm k} - {\bm k}' \right)^2\left({\bm k}' - \frac{\bm\Delta}{2} \right)^2}, 
\end{eqnarray}
and in the third term, ${\cal A}_3(\bm k,\bm \Delta)$, 
\begin{eqnarray}
d^2{\bm k}'\omega_3  = d^2{\bm k'}\frac{\left({\bm k} - \left(-\frac{{\bm \Delta}}{2}\right)\right)^2}{\left({\bm k} -{\bm k'} \right)^2\left({\bm k'} - \left(-\frac{{ {\bm \Delta}}}{2}\right) \right)^2},  
\end{eqnarray}

both are $PSL(2, \mathbb{C})$ invariant. There is, however, a subtle difference.
While $d^2\bm k'\omega_2$ is invariant under the linear fractional transformation on $\left({\bm k},{\bm k'}, \frac{\bm \Delta}{2} \right)$, the measure-kernel combination in the third term, $d^2{\bm k'}\omega_3$, is invariant when  the linear fractional transformation is applied to $\left({\bm k}, {\bm k'}, -\frac{\bm\Delta}{2} \right)$. 
%
%
%

\item The kernel in the non-linear fourth term ${\cal A}_4(\bm k,\bm \Delta)$ is unity $i.e.$ $\omega_4 =1$, leaving only the measure $d^2{\bm \Delta'}$, which is invariant under translation only $({\bm \Delta'} \rightarrow {\bm \Delta}' + \beta)$. This non-linear term is, in fact, a convolution integral over $\bm\Delta'$. The term is symmetric in ${\bm \Delta} \rightarrow - {\bm \Delta}$.
\item In principle, there should be a non-linear odderon-odderon term ${\cal A}_5(\bm k,\bm \Delta)\sim {\cal O}{\cal O}$ in Eq.\eqref{equa01} \cite{Hatta:2005as}. This term is subleading, and we have not explicitly shown it in Eq.\eqref{equa01} for brevity. The term, however, represents the interesting case of merging two $C$-odd odderons into one pomeron. ${\cal A}_5$ is therefore even in ${\bm \Delta} \rightarrow - {\bm \Delta}$. All subsequent discussions will be equally valid for this term as well.
\end{itemize} 

\noindent  To summarise, the five terms in Eq.\eqref{equa01} transform as follows, 
\begin{eqnarray}
{\cal A}_{1,4,5}({\bm k},-{\bm\Delta}) &\rightarrow& {\cal A}_{1,4,5}({\bm k},{\bm\Delta}), \\
{\cal A}_{2,3}({\bm k},-{\bm\Delta}) &\rightarrow& {\cal A}_{3,2}({\bm k},{\bm\Delta}), 
\end{eqnarray}
leaving $\partial_Y{\cal N}({\bm k}, {\bm\Delta})$ invariant under ${\bm \Delta} \rightarrow - {\bm \Delta}$. 

\vspace{0.5cm}
{\color{Red} $\bullet$} {\color{NavyBlue} {\it Exploiting the translation symmetry:~}}
\noindent The discussion above establishes the fact that the terms in the right of Eq.\eqref{equa01} can be grouped into two as,

\begin{eqnarray}
\frac{\partial}{\partial Y}{\cal N}({\bm k}, {\bm \Delta}) ={\cal G}({\bm k}, {\bm \Delta}) + {\cal G}({\bm k}, -{\bm \Delta})
\end{eqnarray}
where, 
\begin{widetext}
\begin{eqnarray}
{\cal G}({\bm k}, {\bm \Delta})&=& \frac{\bar{\alpha}_s}{2\pi} \int d^2 {\bm k}' \left \{\frac{{\cal N}({\bm k}', {\bm \Delta})}{\left({\bm k} - {\bm k}' \right)^2} -\frac{1}{2} \frac{\left({\bm k}-\frac{{\bm \Delta}}{2}\right)^2}{\left({\bm k} - {\bm k}' \right)^2\left({\bm k}'-\frac{{\bm \Delta}}{2}\right)^2}~{\cal N}({\bm k}, {\bm \Delta}) - \frac{1}{2} {\cal N}\left({\bm k} + \frac{{\bm k}'}{2}, {\bm \Delta} - {\bm k}'\right)
{\cal N}\left({\bm k} + \frac{{\bm k}'-{\bm \Delta}}{2}, {\bm k}'\right)\right \}.     \nonumber \\ \label{g10}
\end{eqnarray}
\end{widetext}
While pomeron amplitude ${\cal N}(\bm k,\bm \Delta)$ is symmetric under the transformation $\bm \Delta \rightarrow -\bm \Delta$, the function ${\cal G}(\bm k,\bm \Delta)$ is not symmetric. Under the transformation $\bm \Delta \rightarrow -\bm \Delta$, ${\cal G}(\bm k,\bm \Delta)$ maps to ${\cal G}(\bm k,-\bm \Delta)$ and vice-versa. However, all the kernels in the right-hand side of Eq.\eqref{g10} have the translation symmetry,
\begin{eqnarray}
{\bm k} \rightarrow {\bm k} - \frac{{\bm a}}{2},~~~ {\bm k'} \rightarrow {\bm k'} - \frac{{\bm a}}{2}, ~~~ {\bm\Delta} \rightarrow {\bm \Delta} - {\bm a}. \label{g11}
\end{eqnarray}
As a consequence of this translation symmetry of the kernels, if the initial conditions are also symmetric, the function ${\cal G}(\bm k,\bm \Delta)$ would preserve these symmetries. For integro-differential equations, such as the BK equation, the evolution depends very weakly on the initial conditions, and even if the translation symmetry is not there at the beginning, the evolution will eventually restore the translation symmetry in ${\cal G}$ leading to,   
\begin{eqnarray}
{\cal G}(\bm k, \bm\Delta) = {\cal G}\left({\bm k} - \frac{{\bm a}}{2},\bm \Delta - {\bm a}\right). \label{g12}
\end{eqnarray} 
Therefore, while ${\cal N}(\bm k,\bm \Delta)$ is symmetric in $\bm \Delta \rightarrow -\bm \Delta$ but not translationally symmetric (as defined in Eq.\eqref{g11}), the function ${\cal G}(\bm k,\bm \Delta)$ is translationally symmetric under the translation in Eq.\eqref{g11}, even though is not symmetric in $\bm\Delta\rightarrow-\bm\Delta$.
Now we take the interesting case of taking $\bm a=\bm \Delta$ in Eq.\eqref{g12}. This amounts to substituting ${\bm k} \rightarrow \left({\bm k}-\frac{{\bm \Delta}}{2}\right)$ and ${{\bm \Delta}\rightarrow{\bm 0}}$ in Eq.\eqref{g10} as, 
\begin{widetext}
\begin{eqnarray}\nonumber
{\cal G}\left({\bm k} - \frac{{\bm \Delta}}{2}, {\bm 0} \right) &=& \frac{\bar{\alpha}_s}{2\pi} \int d^2 {\bm k}' \left \{\frac{{\cal N}\left({\bm k}', {\bm 0}\right)}{\left(\left({\bm k} - \frac{{\bm \Delta}}{2}\right) - {\bm k}' \right)^2} -\frac{1}{2} \frac{\left({\bm k}-\frac{{\bm \Delta}}{2}\right)^2}{\left(\left({\bm k} - \frac{{\bm \Delta}}{2}\right) - {\bm k}' \right)^2{\bm k}'^2}~{\cal N}\left({\bm k-\frac{\Delta}{2}}, {\bm 0}\right)\right. \\
&&~~~~~~~~~~~~~~~~~~~~~~~~~~~~~~~~~~~~~~~~~~~~~~~ \left.- \frac{1}{2} {\cal N}\left({\bm k} - \frac{{\bm \Delta}}{2} + \frac{{\bm k}'}{2},  - {\bm k}'\right)
{\cal N}\left({\bm k} - \frac{{\bm \Delta}}{2} + \frac{{\bm k}'}{2}, {\bm k}'\right)\right \}. \label{g13}    
\end{eqnarray}
\end{widetext}
Note that $\bm k'$, being a dummy variable, is integrated over. Interestingly, the right hand side of Eq.\eqref{g13} can be obtained by substituting ${\bm k} \rightarrow \left({\bm k}-\frac{{\bm \Delta}}{2}\right)$ and ${{\bm \Delta}\rightarrow{\bm 0}}$ in Eq.\eqref{equa01}. Therefore, 
\begin{eqnarray}
{\cal G}\left({\bm k} - \frac{{\bm \Delta}}{2}, {\bm 0} \right) &=&\frac{1}{2}\frac{\partial}{\partial Y}  {\cal N}\left(\bm k-\frac{\bm\Delta}{2},{\bm 0}\right). 
\end{eqnarray}
\noindent With this, the Eq.\eqref{equa01} can be rewritten in the following way, 
\begin{eqnarray} \nonumber 
&&\frac{\partial}{\partial Y} {\cal N}({\bm k}, {\bm \Delta}) ~~~~~~~\\
&&=\frac{1}{2} \frac{\partial}{\partial Y}\left[ {\cal N}\left({\bm k} - \frac{{\bm \Delta}}{2},~{\bm 0}\right) + {\cal N}\left({\bm k} + \frac{{\bm \Delta}}{2},~{\bm 0}\right) \right].\nonumber \\ \label{equariya}
\end{eqnarray}
This is the main result of this article. It shows that the small-$x$ evolution of non-forward pomeron amplitude ${\cal N}({\bm k}, {\bm \Delta})$, within the leading-log approximation, is completely determined by the small-$x$ evolution of forward amplitudes ${\cal N}\left({\bm k} - \frac{\bm \Delta}{2}, {\bm 0}\right)$ and ${\cal N}\left({\bm k} +\frac{\bm \Delta}{2}, {\bm 0}\right)$.\\

\vspace{0.5cm}
\noindent {\color{Red} $\bullet$} {\color{NavyBlue} {\it Evolution of Odderon amplitude:~}} 
Recently, the TOTEM collaboration at the LHC and the DØ collaboration at the former Tevatron collider at Fermilab, jointly announced the discovery of the odderon in $pp$ and $p\bar p$  scattering at non-zero momentum transfer \cite{D0:2020tig}. Odderon is an elusive three-gluon state. While pomerons are made of two reggeized gluons in the colorless state, the odderons are objects made of three reggeized gluons in their $d$-color state. It has been found that the high energy intercept for odderon is exactly one. 
A similar analysis, as discussed before for pomeron,  can be extended for Odderon amplitude as well. While the Fourier transform would be identical to Pomeron, being the $C$-odd object ${\cal O}({\bm k}, {\bm \Delta})$ would transform as 
${\cal O}({\bm k}, {\bm \Delta})= -{\cal O}({\bm k}, {-\bm \Delta})$ when one flips the sign of $\bm\Delta$. This will ensure that $\partial{\cal O}({\bm k}, {\bm \Delta})/\partial Y $ can be expressed as the \textit{difference} of two translationally invariant functions, contrary to the pomeron amplitude, as presented in Eq.\eqref{equariya} where, $\partial{\cal N}({\bm k}, {\bm \Delta})/\partial Y $ is the \textit{sum} of two translationally invariant functions ${\cal G}({\bm k}, {\bm \Delta}) $ and ${\cal G}({\bm k}, {-\bm \Delta}) $ (details are in the appendix).
This leads to the following expression for the evolution of odderon amplitude at high energy at leading-log approximation, 
\begin{eqnarray}\nonumber
&&\frac{\partial}{\partial Y} {\cal O}({\bm k}, {\bm \Delta})~~~~~~~ \\
&&=\frac{1}{2} ~ \frac{\partial}{\partial Y}  \left[ {\cal O}\left({\bm k} - \frac{{\bm \Delta}}{2},{\bm 0}\right) - {\cal O}\left({\bm k} + \frac{{\bm \Delta}}{2},{\bm 0}\right) \right]. \nonumber \\  \label{equariya2} 
\end{eqnarray}
It is evident from Eq.\eqref{equariya2} that the odderon amplitude does not evolve with energy in the forward limit, $\bm\Delta=0$, corroborating the fact that odderons are more likely to be found in the non-forward scatterings.
%
%

%
%

\vspace{0.5cm}
\noindent {\color{Red} $\bullet$} {\color{NavyBlue} {\it Estimation of pomeron and odderon amplitudes at non-zero $\bm\Delta$}:~}\\
Small-$x$ evolution for pomeron and odderon amlitudes as presented in Eq.\eqref{equariya} and Eq.\eqref{equariya2} are the key results of this work. In the following we will use these two equations to estimate the pomeron and odderon amplitudes at $\bm\Delta\neq0$.
\begin{itemize}
    \item {\it Region of linear evolution $\bm k \approx Q_s$}:~

%
%
In the vicinity of the saturation boundary, when the non-linear terms can be dropped in the evolution Eq.\eqref{equariya}, the pomeron amplitude at forward limit can be written as, ${\cal N}(\bm k)\sim \left({Q^2_{s}(Y)}/{{\bm k}^2}\right)^{\gamma_{cr}}$, where $Q^2_{s}(Y)$ is the saturation scale and $\gamma_{cr}\approx 0.63$. 
This expression for the amplitude at forward limit can be substituted in Eq.\eqref{equariya} to get the amplitude at non-zero $\bm\Delta$ as,
\begin{eqnarray}
   {\cal N}\left({\bm k},{\bm \Delta}\right)&\sim &\frac{1}{2} \left[\left(\frac{Q^2_{s}(Y)}{\left({\bm k}-\frac{{\bm \Delta}}{2}\right)^2}\right)^{\gamma_{cr}}+\left(\frac{Q^2_{s}(Y)}{\left({\bm k} + \frac{{\bm \Delta}}{2}\right)^2}\right)^{\gamma_{cr}}\right]. \nonumber \\
\end{eqnarray}
The Odderon, however, has additional suppression factor in high-energy as ${\cal O}(\bm k)\sim \left({Q^2_{s}(Y)}/{{\bm k}^2}\right)^{\gamma_{0}}e^{-\tau_0 \alpha_s Y}$ where $\gamma_0 \sim 1.04$ and $\tau_0 \sim 4.11$ \cite{Motyka:2005ep}. The odderon evolution equation in Eq.\eqref{equariya2} then leads to the following expression in the dilute regime at non-zero $\bm\Delta$ as,
\begin{eqnarray} \nonumber
   {\cal O}\left({\bm k}, {\bm \Delta}\right)& \sim &\frac{1}{2} \left[\left(\frac{Q^2_{s}(Y)}{\left({\bm k} - \frac{{\bm \Delta}}{2}\right)^2}\right)^{\gamma_{0}}-\left(\frac{Q^2_{s}(Y)}{\left({\bm k} + \frac{{\bm \Delta}}{2}\right)^2}\right)^{\gamma_{0}}\right]\\
&&~~~~~~~~~~~~~~~~~~~~~~~~~~~~\times~ e^{-\tau_0 \alpha_s Y}.  
\end{eqnarray} 
\item {\it Region of non-linear evolution $\bm k \ll Q_s$}:~ Deep inside the saturation boundary, where non-linear evolution is important, the pomeron amplitude behaves as, ${\cal N}(\bm k)\sim \ln\left({Q^2_{s}(Y)}/{{\bm k}^2}\right)$ \cite{Siddiqah:2018qey,Amaral:2020xqv}, leading to 
\begin{eqnarray}
     {\cal N}\left({\bm k}, {\bm \Delta}\right) \sim \ln\left(\frac{Q^2_{s}(Y)}{\left|{\bm k} - \frac{{\bm \Delta}}{2}\right|\left|{\bm k} + \frac{{\bm \Delta}}{2}\right|}\right). 
\end{eqnarray}

\end{itemize}


\vspace{0.5cm}
\noindent {\color{Red} $\bullet$} {\color{NavyBlue} {\it Estimations of unpolarised dipole GTMD $f_{1,1}$:}~}
While the efforts to understand QCD at small-$x$ have a steady chronicle of progress, the recent initiatives around the installation of the upcoming Electron-Ion Collider (EIC) have triggered a surge in activity towards understanding the small-$x$ QCD evolution, proton's 3D structure, and associated spin correlations. The study of proton's 3D structure in momentum space often revolves around understanding the transverse momentum-dependent parton distribution functions (or TMDs) and Generalized TMDs (GTMDs) (both polarised and unpolarised) and their evolution across the scales and energy.
While TMDs are functions of Bjorken-$x$ variable and the transverse momentum ${\bm k}$, the GTMDs additionally contain the momentum transfer ${\bm \Delta}$.  Both TMDs and GTMDs are non-perturbative objects; hence, they can be probed only through experiments. However, their evolution along scales and energy can be systematically estimated within the perturbative framework.  
The GTMDs encode generalized parton distributions (GPDs) and the transverse momentum-dependent parton distributions (TMDs) in themselves and are often attributed as the parent distribution. The GTMDs are related to Wigner distributions through Fourier transform to the transverse position space at zero skewness parameter.
Specific gluon GTMD $F_{1,4}$ is also directly connected to the gluon Orbital Angular Momentum (OAM) - and is often seen as an avenue to probe the gluon OAM at the upcoming collider \cite{Lorce:2011kd,Hatta:2011ku,Lorce:2011ni}. 
%
Being relatively new, GTMD studies are still limited to either model calculations or near-forward limits (the small $\bm \Delta$ approximation). 
The unpolarised color TMDs and GTMDs are directly connected to the dipole amplitudes. In fact the unpolarised colored GTMD is also governed by the identical equation as presented in Eq.\eqref{equa01} for pomeron amplitude even in the presence of spin. The parametrization of ${\cal N}(\bm k,\bm\Delta)$ would still be valid after taking its matrix elements between high energy proton states with definite spin states for transversely polarized proton. The amplitude ${\cal N}(\bm k,\bm\Delta)$ can be paramterized in terms of various GTMDs as \cite{Hatta:2022bxn},
\begin{eqnarray} \nonumber
    {\cal N}(\bm k,\bm\Delta,\bm S)&\approx& \frac{\pi g^2}{2N_c}\left\{{\cal F}_{1,1}(\bm k,\bm\Delta)-i\bm k \times \bm S\frac{\bm k.\bm\Delta}{M^4}{\cal F}_{1,2}(\bm k,\bm\Delta)\right.\\
  &&\left. -i\frac{\bm \Delta\times \bm S}{2M^2}\left[2{\cal F}_{1,3}(\bm k,\bm\Delta)-{\cal F}_{1,1}(\bm k,\bm\Delta)\right]\right\}.\label{eq29}
\end{eqnarray}
One can therefore write similar relations, for various GTMDs, that connect them with TMDs as we have done for pomeron and odderon amplitudes in Eq. \eqref{equariya} and Eq.\eqref{equariya2}. For example, the function ${\cal F}_{1,1}$, as defined in Eq.\eqref{eq29}, also satisfy Eq.\eqref{equa01} and one can write,
\begin{eqnarray}\nonumber
&& \frac{\partial}{\partial Y} {\cal F}_{1,1}\left(\bm k,\bm \Delta\right)~~~\\
 &&  = \frac{1}{2}\frac{\partial}{\partial Y}\left[{\cal F}_{1,1}\left(\bm k-\frac{\bm \Delta}{2},0\right)+{\cal F}_{1,1}\left(\bm k+\frac{\bm \Delta}{2},0\right)\right].\nonumber\\
\end{eqnarray}
The function ${\cal F}_{1,1}$ is directly connected to unpolarized gluon GTMD $f_{1,1}$ as (pomeron part)\cite{Hatta:2022bxn},
\begin{eqnarray}
    f_{1,1}(\bm k,\bm\Delta)&=&{\bm k^2} \frac{\partial^2}{\partial \bm k^\alpha\partial \bm k^\alpha} {\cal F}_{1,1}(\bm k,\bm\Delta).
\end{eqnarray}
In recent years, there have been extensive studies taken place on the small-$x$ evolution of various TMDs or related non-perturbative objects \cite{Abir:2015qva,Vasim:2019bdf,Hatta:2022bxn,Bhattacharya:2023yvo,Agrawal:2023mzm}. This connection of GTMDs with TMDs may help in estimating GTMDs through their TMD counter-parts and go beyond small $\bm \Delta$ approximation.


\vspace{0.5cm}
\noindent {\color{Red} $\bullet$} {\color{NavyBlue} {\it Conclusion:}~}
In this paper, we have shown that the evolution of the off-forward pomeron amplitude in the $\bm k$-space can be studied through the evolution of the sum of two bifurcated forward pomeron amplitudes. The same arguments has been extended to the odderon amplitude and it is shown that it's evolution can be studied through the evolution of the difference of the forward odderon amplitudes. This result can have further applications, extending to GTMDs and TMDs including the polarized ones.
The phenomenological consequences of the off-forward limit, $\bm\Delta\neq0$, are often neglected for convenience. With the upcoming experiments like EIC and the recent theoretical efforts to understand spin orbit correlation/entanglement \cite{Bhattacharya:2024sck,Bhattacharya:2024sno,Hatta:2024otc,Hatta:2024lbw}, it has become more important to study the high-energy evolutions with non-zero momentum transfer. The study of GTMDs has also gained much popularity since it has been shown to be obtained from diffractive di-jet production in the deep-inelastic electron-proton and electron-ion collisions \cite{Hatta:2016dxp, Hatta:2016aoc, Ji:2016jgn,Bhattacharya:2023hbq}.
The result presented in this paper may open new ways to access GTMDs via TMDs.

\newpage
\begin{widetext}
\section{Appendix}

\subsection{Fourier transformation of real symmetric functions in two-dimension} 

\noindent For the convenience of the discussion that follows, let's take a real function ${F}({\bm x}, {\bm y})$ of two position vectors ${\bm x}$, ${\bm y}$ in a plane. The function ${F}({\bm x}, {\bm y})$ is symmetric under the exchange: ${\bm x} \leftrightarrow  {\bm y}$. We are interested in the Fourier transformation of this function in momentum space. If ${\bm k_x}$ and ${\bm k_y}$ are momentum conjugate to ${\bm x}$, ${\bm y}$ respectively, the Fourier transformed function can be written as,  
\begin{eqnarray}
 {\cal F} ({\bm k_x}, {\bm k_y}) = \int \frac{d^2{\bm x}}{2\pi} \frac{d^2{\bm y}}{2\pi} ~ e^{i {\bm k_x}{\bm x}} e^{i {\bm k_y} {\bm y}} ~  
 F({\bm x}, {\bm y}).
\end{eqnarray}
It is evident that if the function, ${F}({\bm x}, {\bm y})$ is symmetric under ${\bm x}\leftrightarrow {\bm y}$, this symmetry also translates to the momentum space $i.e.$ ${\cal F}({\bm k_x}, {\bm k_y})$ is also invariant under the exchange ${\bm k_x} \leftrightarrow {\bm k_y}$. 
We now transform the function ${F}({\bm x}, {\bm y})$ to ${\bar{F}}({\bm r}, {\bm b})$ where $({\bm r}, {\bm b})$ are defined, as usual, 
\begin{eqnarray}
{\bm r} &=& \left( {\bm x} - {\bm y} \right), \nonumber \\
{\bm b} &=& \left({\bm x} + {\bm y}\right)/2.
\end{eqnarray}
In the literature on the scattering of particles, ${\bm r}$ is identified as the transverse separation, and ${\bm b}$ is defined as the impact parameter. 
\noindent Now lets Fourier transform the function ${\bar F}({\bm r}, {\bm b})$ to ${\bar {\cal F}({\bm k}}, {\bm \Delta})$, where ${\bm k}$ is conjugate momentum to ${\bm r}$ and ${\bm \Delta}$ is conjugate to ${\bm b}$, 
\begin{eqnarray}
{\bar {\cal F}} ({\bm k}, {\bm \Delta}) = \int \frac{d^2{\bm r}}{2\pi} \frac{d^2{\bm b}}{2\pi} ~ e^{i {\bm k}{\bm r}} e^{i {\bm \Delta} {\bm b}} ~  {\bar F}({\bm r}, {\bm b}).\label{qq1}
\end{eqnarray}
As the function ${\bar F}({\bm r}, {\bm b})$ is not generally symmetric under ${\bm r}\leftrightarrow {\bm b}$, the Fourier transformed function ${\bar {\cal F}} ({\bm k}, {\bm \Delta})$ is also expected not to be symmetric under the exchange ${\bm k}\leftrightarrow {\bm \Delta}$.
As the Jacobian of transformation is unity, when expressed in $({\bm x}, {\bm y})$, the Eq.\eqref{qq1} can be written as, 
\begin{eqnarray}
{\bar {\cal F}} ({\bm k}, {\bm \Delta}) = \int \frac{d^2{\bm x}}{2\pi} \frac{d^2{\bm y}}{2\pi} ~
\exp\left[i\left({\bm k} + \frac{\bm \Delta}{2}\right){\bm x} + i\left({\bm k} - \frac{\bm \Delta}{2}\right){\bm y}\right] ~  F({\bm x}, {\bm y}).
\end{eqnarray}
As $F({\bm x},{\bm y})$ is real and symmetric under the exchange of ${\bm x} \leftrightarrow {\bm y}$, following symmetry properties hold,  
\begin{eqnarray}
&& {\it Re} ~ {\cal F}({\bm k},{\bm \Delta}) = {\it Re} ~ {\cal F}(-{\bm k},{\bm \Delta}) = {\it Re} ~ {\cal F}({\bm k},-{\bm \Delta})
=  {\it Re} ~ {\cal F}(-{\bm k},-{\bm \Delta}), \\
&& {\it Im} ~ {\cal F}({\bm k},{\bm \Delta}) = {\it Im} ~ {\cal F}(-{\bm k},{\bm \Delta}) = - {\it Im} ~ {\cal F}({\bm k},-{\bm \Delta})
=  - {\it Im} ~ {\cal F}(-{\bm k},-{\bm \Delta}).
\end{eqnarray}
Under the exchange $\left({\bm k} - \frac{\bm \Delta}{2}\right) \leftrightarrow \left({\bm k} + \frac{\bm \Delta}{2}\right)$, while ${\it Re} ~ {\cal F}({\bm k},{\bm \Delta})$ is symmetric, ${\it Im} ~ {\cal F}({\bm k},{\bm \Delta})$ is anti-symmetric. 
Transformations such as $({\bm x},{\bm y})\rightarrow ({\bm r},{\bm b})$ or $\left({\bm k},{\bm \Delta}\right) \rightarrow \left( {\bm k-\frac{\bm \Delta}{2}},{\bm k+\frac{\bm \Delta}{2}}\right)$ are not linear transformations. Rather, they belong to broader affine transformations that involve linear transformation along with a non-linear translation of coordinates. However, 
in the normed vector spaces where $({\bm x},{\bm y})$ or $({\bm k},{\bm \Delta})$ belong and inner products are defined, the statement of the parallelogram law is essentially equations that relate their norms, 
\begin{eqnarray}
 {\bm x}^2 + {\bm y}^2 &=& \frac{{\bm r}^2}{2} +  2{\bm b}^2 = \left({\bm b} - \frac{\bm r}{2}\right)^2 + \left( {\bm b} + \frac{\bm r}{2} \right)^2,  \nonumber \\
\left({\bm k} - \frac{\bm \Delta}{2}\right)^2 + \left( {\bm k} + \frac{\bm \Delta}{2} \right)^2 &=&
 2{\bm k}^2 + \frac{{\bm \Delta}^2}{2}.
\end{eqnarray}
When $F({\bm x},{\bm y})$ has additional symmetry: $F(-{\bm x},~-{\bm y}) = F({\bm x},~{\bm y})$, then ${\it Im} ~ {\cal F}({\bm k},{\bm \Delta})=0$ $i.e.$ ${\cal F}({\bm k},{\bm \Delta})$ is real.

\subsection{The first term} 
\noindent {\color{NavyBlue} $\bullet$} The first term ${\cal A}_1$ is symmetric in $\bm\Delta$ as, 
\begin{eqnarray}
{\cal A}_1(\bm k,-\bm\Delta) = \int \frac{d^2 \bm k'}{\left(\bm k' -\bm k \right)^2} ~ {\cal N}(\bm k', -\bm\Delta) = \int \frac{d^2 \bm k'}{\left(\bm k' -\bm k \right)^2} ~ {\cal N}(\bm k', \bm\Delta) = {\cal A}_1(\bm k,\bm\Delta).  
\end{eqnarray}

\subsection{The non-linear fourth term}

\noindent {\color{NavyBlue} $\bullet$} The non-linear fourth term is also symmetric in $\bm\Delta$ as, 
\begin{eqnarray}
{\cal A}_4(\bm k,\bm\Delta) &=& \int d^2 \bm\Delta' d^2 \bm\Delta'' {\cal N}\left(\bm k -\frac{\bm\Delta}{2}+ \frac{\bm\Delta'}{2}, \bm\Delta'\right)
{\cal N}\left(\bm k + \frac{\bm\Delta}{2} -\frac{\bm\Delta''}{2}, \bm\Delta''\right)\delta^2(\bm\Delta'+\bm\Delta''-\bm\Delta), \nonumber \\
&=& \int d^2 \bm\Delta' d^2 \bm\Delta'' {\cal N}\left(\bm k - \frac{\bm\Delta'}{2}, -\bm\Delta''\right)
{\cal N}\left(\bm k + \frac{\bm\Delta''}{2}, -\bm\Delta'\right)\delta^2(-\bm\Delta'-\bm\Delta''+\bm\Delta),  \nonumber \\
&=& \int d^2 \bm\Delta' d^2 \bm\Delta'' {\cal N}\left(\bm k - \frac{\bm\Delta'}{2}, \bm\Delta''\right)
{\cal N}\left(\bm k + \frac{\bm\Delta''}{2}, \bm\Delta'\right)\delta^2(\bm\Delta'+\bm\Delta''-\bm\Delta),  \nonumber \\
&=& {\cal A}_4(\bm k,-\bm\Delta), 
\end{eqnarray}
where in the second line, we have flipped the sign of $\bm\Delta'$ and $\bm\Delta''$ as the integrations are in the symmetric interval.

\subsection{Evolution of Odderon} 
\noindent The evolution equation for Odderon in the momentum space is,
\begin{eqnarray} \nonumber
 \frac{\partial}{\partial Y} {\cal O}({\bm k}, {\bm \Delta})&=& \frac{\bar{\alpha}_s}{\pi} \int \frac{d^2 {\bm k}'}{\left({\bm k}' - {\bm k} \right)^2} \left \{{\cal O}({\bm k}', {\bm \Delta}) -\frac{1}{4} \frac{\left({\bm k}-\frac{{\bm \Delta}}{2}\right)^2}{\left({\bm k}'-\frac{{\bm \Delta}}{2}\right)^2}~{\cal O}({\bm k}, {\bm \Delta})-\frac{1}{4}\frac{\left({\bm k}+\frac{{\bm \Delta}}{2}\right)^2}{\left({\bm k}'+\frac{{\bm \Delta}}{2}\right)^2}~{\cal O}({\bm k}, {\bm \Delta}) \right \} \\ \nonumber  
&&-\frac{\bar{\alpha}_s}{2\pi}\int d^2{\bm \Delta}' \left[{\cal O}\left({\bm k} + \frac{{\bm \Delta}'}{2}, {\bm \Delta} - {\bm \Delta}'\right)
{\cal N}\left({\bm k} + \frac{{\bm \Delta}'-{\bm \Delta}}{2}, {\bm \Delta}'\right)+{\cal N}\left({\bm k} + \frac{{\bm \Delta}'}{2}, {\bm \Delta} - {\bm \Delta}'\right)
{\cal O}\left({\bm k} + \frac{{\bm \Delta}'-{\bm \Delta}}{2}, {\bm \Delta}'\right)\right]   \\ \label{Odd10}
\end{eqnarray}
Unlike pomeron ${\cal N}(\bm k,\bm \Delta)$, the odderon amplitude ${\cal O}(\bm k,\bm \Delta)$ is anti-symmetric under the transformation, $\bm \Delta\rightarrow-\bm\Delta$,
\begin{eqnarray}
    {\cal O}(\bm k,\bm \Delta)&=&- ~ {\cal O}(\bm k,-\bm \Delta).
\end{eqnarray}
Thus, we can group the right side of the evolution equation for Odderon in Eq.\eqref{Odd10} into two,
\begin{eqnarray}
    \partial_Y {\cal O}({\bm k}, {\bm \Delta})&=&{\cal J}(\bm k,\bm \Delta)-{\cal J}(\bm k,-\bm \Delta), \label{odd3}
\end{eqnarray}
where,
\begin{eqnarray}
    \frac{\partial}{\partial Y} {\cal J}({\bm k}, {\bm \Delta})&=& \frac{\bar{\alpha}_s}{2\pi} \int d^2 {\bm k}' \left [\frac{{\cal O}({\bm k}', {\bm \Delta})}{\left({\bm k}' - {\bm k} \right)^2} -\frac{1}{2} \frac{\left({\bm k}-\frac{{\bm \Delta}}{2}\right)^2}{\left({\bm k}' - {\bm k} \right)^2\left({\bm k}'-\frac{{\bm \Delta}}{2}\right)^2}~{\cal O}({\bm k}, {\bm \Delta})\right. \nonumber  \\ 
    &&~~~~\left.-\frac{1}{2} \left\{{\cal O}\left({\bm k} + \frac{{\bm k}'}{2}, {\bm \Delta} - {\bm k}'\right)
{\cal N}\left({\bm k} + \frac{{\bm k}'-{\bm \Delta}}{2}, {\bm k}'\right)+{\cal N}\left({\bm k} + \frac{{\bm k}'}{2}, {\bm \Delta} - {\bm k}'\right){\cal O}\left({\bm k} + \frac{{\bm k}'-{\bm \Delta}}{2}, {\bm k}'\right)\right\} \right].  \nonumber \\ \label{qq}
\end{eqnarray}
 As the case for pomeron, all kernels in Eq.\eqref{qq} are invariant under the translation,
\begin{eqnarray}
    \bm k \rightarrow \bm k-\frac{\bm a}{2}, ~~~ {\bm k}' \rightarrow {\bm k}' -\frac{\bm a}{2} ~~ \text{and}~~~ \bm \Delta\rightarrow\bm \Delta-\bm a.
\end{eqnarray}
Hence, ${\cal J}({\bm k}, {\bm \Delta})$ possesses the following symmetry, 
\begin{eqnarray}
{\cal J}(\bm k, \bm\Delta) = {\cal J}\left({\bm k} - \frac{{\bm a}}{2},\bm \Delta - {\bm a}\right).
\end{eqnarray} 
Putting $\bm a=\bm \Delta$ in Eq.\eqref{qq}, we get
\begin{eqnarray}
     \frac{\partial}{\partial Y} {\cal J}({\bm k}-\frac{\bm \Delta}{2}, \bm 0)&=& \frac{\bar{\alpha}_s}{2\pi} \int d^2 {\bm k}' \left [\frac{{\cal O}({\bm k}'-\frac{\bm \Delta}{2}, {\bm 0})}{\left({\bm k}' - {\bm k} \right)^2} -\frac{1}{2} \frac{\left({\bm k}-\frac{{\bm \Delta}}{2}\right)^2}{\left({\bm k}-\frac{\bm \Delta}{2} - {\bm k}' \right)^2\left({\bm k}'-\frac{{\bm \Delta}}{2}\right)^2}~{\cal O}({\bm k}-\frac{\bm \Delta}{2}, {\bm 0})\right. \nonumber \\ 
    && \left.-\frac{1}{2} \left\{{\cal O}\left({\bm k}-\frac{\bm \Delta}{2} + \frac{{\bm k}'}{2},  - {\bm k}'\right)
{\cal N}\left({\bm k}-\frac{\bm \Delta}{2} + \frac{{\bm k}'}{2}, {\bm k}'\right)+{\cal N}\left({\bm k}-\frac{\bm \Delta}{2} + \frac{{\bm k}'}{2}, - {\bm k}'\right){\cal O}\left({\bm k}-\frac{\bm \Delta}{2} + \frac{{\bm k}'}{2}, {\bm k}'\right)\right\} \right] \nonumber \\
    &=&\frac{1}{2}\frac{\partial}{\partial Y} {\cal O}\left(\bm k-\frac{\bm\Delta}{2}, {\bm 0} \right).
\end{eqnarray}
Thus, using \eqref{odd3} we can write the evolution of ${\cal O}({\bm k},{\bm \Delta})$ as,
\begin{eqnarray}
    \frac{\partial}{\partial Y} {\cal O}({\bm k},{\bm \Delta})&=&
    \frac{1}{2}~\frac{\partial}{\partial Y}\left[ {\cal O}({\bm k}-\frac{\bm \Delta}{2}, \bm 0)- {\cal O}({\bm k}+\frac{\bm \Delta}{2}, \bm 0)\right].
\end{eqnarray}

\end{widetext}

\bibliography{ref.bib}

\end{document}